\documentclass[twocolumn,showpacs,preprintnumbers,amsmath,amssymb]{revtex4}
\usepackage{epsfig}
\usepackage{graphics}
\usepackage{graphicx}% Include figure files
\usepackage{dcolumn}% Align table columns on decimal point
\usepackage{bm}% bold math

\begin{document}

%\preprint{APS/123-QED}

\title{Global phase synchronization in an array of time-delay systems}% Force line breaks with \\

\author{R.~Suresh$^1$}
\email{suresh@cnld.bdu.ac.in}
\author{D.~V.~Senthilkumar$^{2,3}$}
\email{skumar@cnld.bdu.ac.in}
\author{M.~Lakshmanan$^1$}%
 \email{lakshman@cnld.bdu.ac.in}
 \author{J.~Kurths$^{3,4}$}
 \email{juergen.kurths@pik-potsdam.de}
\affiliation{
$^1$Centre for Nonlinear Dynamics, Department of Physics, Bharathidasan University, Tiruchirapalli-620024, India\\
$^2$Centre for Dynamics of Complex Systems, University of Potsdam, 14469 Potsdam, Germany\\
$^3$Potsdam Institute for Climate Impact Research, 14473 Potsdam, Germany\\
$^4$Institute of Physics, Humboldt University, 12489 Berlin, Germany}

\date{\today}% It is always \today, today,

\begin{abstract}
	
We report the identification of global phase synchronization (GPS) in a linear array of 
unidirectionally coupled Mackey-Glass time-delay systems exhibiting highly
non-phase-coherent chaotic attractors with complex topological structure.
In particular, we show that the dynamical organization of all the coupled
time-delay systems in the array to form GPS is achieved by sequential 
synchronization as a function of
the coupling strength. Further, the asynchronous ones
in the array with respect to the main sequentially synchronized cluster  
organize themselves to form clusters before they achieve synchronization 
with the main cluster. We have confirmed these
results by estimating instantaneous phases including phase difference,
average phase, average frequency,  frequency ratio and their differences
from suitably transformed phase coherent attractors
after using a nonlinear transformation of the original non-phase-coherent
attractors. The results are further corroborated using two other independent
approaches based on recurrence analysis and the concept of localized
sets from the original non-phase-coherent attractors directly without explicitly
introducing the measure of phase.

\end{abstract}

\maketitle

\section{\label{sec:level1}Introduction}

Chaotic phase synchronization (CPS), referred to as the locking of the phases 
of the coupled chaotically evolving dynamical systems, has been investigated in ensembles of globally 
coupled arrays~\cite {aspmgr2001,sbjk2002,jk2000,mvigvo2004,attf2000,
aspmgr1996,izkyz2002,czjk2002,gvoasp1997,mzzz2000,gkagv2000,kork2000},
networks of oscillators~\cite{sbvl2006,aaadl2008,casbamb2007,qrjz2007} with applications to 
electrochemistry~\cite{izkyz2002,czjk2002}, laser systems~\cite{gkagv2000,kork2000},
cardiorespiratory systems~\cite{csmgr1998,ashh2000,rbjwk2007}, 
neuroscience~\cite{fvjpl2001,jljs2004,ptmgr1998}, 
ecology~\cite{bbah1990,es1990,reagr2006}, climatology~\cite{drsh2003,kyag2009,dmjk2005}, etc. 
Even though the notion of CPS is well studied in low dimensional systems, there exist very little 
indepth studies in higher dimensional systems such as time-delay systems 
which are essentially infinite-dimensional in nature and often exhibit
high-dimensional, highly non-phase-coherent hyperchaotic attractors with 
complex topological structure. Consequently, estimating phase explicitly
to identify phase synchronization in such systems is quite difficult. Nevertheless,
CPS  has been recently demonstrated in two coupled piecewise linear and Mackey-Glass 
time-delay systems~\cite{dvskml2006,dvskml2008} by introducing a nonlinear
transformation of the original dynamical variable to recast the original
non-phase-coherent hyperchaotic attractors into smeared limit cycle-like attractors 
in order to facilitate the estimation of the phase variable using the available methods.
However, these investigations are carried out so far only in two coupled time-delay
systems. In view of the widespread applications of CPS in ensembles of coupled
oscillators, here we investigate the existence of global phase synchronization (GPS)
in an array of unidirectionally coupled Mackey-Glass time-delay 
systems and analyse the mechanism behind the dynamical organization of the coupled
oscillators to form GPS. At first, we use 
the nonlinear transformation introduced in~\cite{dvskml2006,dvskml2008}  to estimate
explicitly the phases of all the oscillators in the array and identify the 
existence of GPS in the array. Further, the existence of GPS is also confirmed 
from the original non-phase-coherent chaotic attractors themselves using
two independent approaches, namely recurrence analysis~\cite{mcrmt2005,nmmcr2007}
and the concept of localized sets~\cite{tpmsb2007}. 

In addition, we will show that the onset of GPS in such arrays does not 
happen instantaneously, but instead takes place as a form of sequential synchronization. 
For lower values of coupling strengths the phases of nearby systems get already synchronized 
with the drive system in contrast to the far away systems. This sequential synchronization of chaotic 
systems can have applications in communication systems~\cite{chil2002}. 
Furthermore, other non-synchronized time-delay systems with
respect to the sequentially synchronized cluster display clusters of phase 
synchronized states among themselves before they become 
synchronized with the large cluster in the sequence to form global phase synchronization.
This clustering is observed when the group of oscillators splits into subgroups such that 
all the oscillators within one cluster move in perfect phase synchrony. This clustering is
considered to be particularly significant in biological systems~\cite{kk1990,as1994,shs1993}. Recently cluster 
synchronization in an array of three chaotic lasers without delay was reported~\cite{jrt1999} as well. 
Also global synchronization via cluster formation has been observed in coupled phase 
oscillators without time-delay~\cite{zheng1998} with simultaneous phase slips of all oscillators, where
quantized phase shifts in these phase slips have been observed. By increasing the coupling, a bifurcation tree
from high-dimensional quasiperiodicity to chaos to quasiperiodicity and 
periodicity has also been found.

The remaining paper is organized as follows: In Sec.~\ref{sec:linear},
we will describe briefly the coupling configuration 
and the nature of chaotic attractors exhibited by the Mackey-Glass time-delay system.
The existence of global phase synchronization 
from the transformed attractors is discussed in Sec.~\ref{sec:pc}, which is
also confirmed from the original non-phase-coherent chaotic attractors 
using recurrence analysis and the concept of localized sets in Sec.~\ref{sec:npc}.
Finally, we summarize our results in Sec.~\ref{sec:con}. 
\begin{figure}
\centering
\includegraphics[width=1.0\columnwidth]{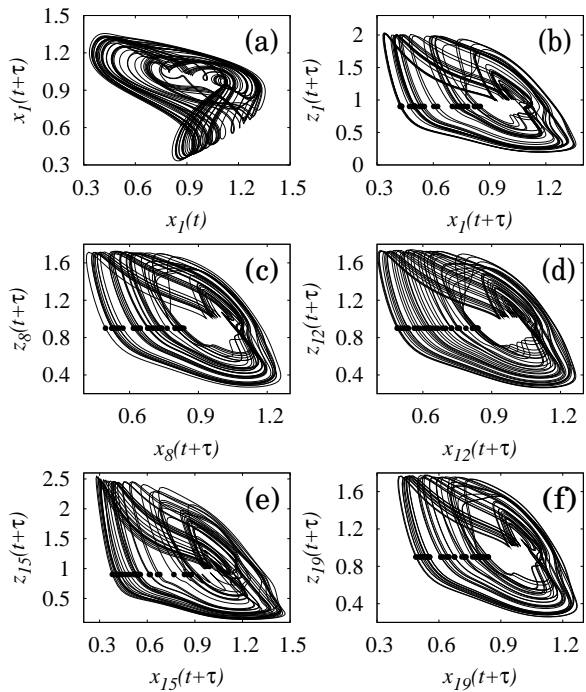}
\caption{\label{fig1} (a) Non phase-coherent chaotic attractor of the drive 
system given by Eq.~(\ref{eq_1}a). (b) Transformed attractor (using Eq.~(\ref{eq_2})) of the drive system. 
(c-f) Transformed attractors of some randomly selected response systems 
($i = 8,12,15,19$) in the projected phase space ($x_i(t+\tau),z_i(t+\tau)$), 
where they look like smeared limit cycle attractors in the absence of the coupling 
along with the Poincar\'e points represented by filled circles. Here $\tau$ has been chosen as 
$\tau$ = 20.0.}
\end{figure}
\section{\label{sec:linear}Linear array of Mackey-Glass time-delay systems}
The Mackey-Glass time-delay system was originally deduced as a 
model for blood production in patients with leukemia~\cite{mcmlg1977},
and it has been well studied in the literature for its hyperchaotic behavior 
~\cite{cschl1999,kp1998,igsjml2005} and has also been experimentally realized 
using analog electronic circuits ~\cite{ankp1995}. In this paper, we consider a linear
array of unidirectionally coupled Mackey-Glass systems with free-end boundary conditions represented by
the following system of coupled nonlinear first order ordinary differential equations, 
\begin{subequations}
\begin{eqnarray}
\dot{x}_1(t)&=&-\beta x_1(t)+\frac{\alpha_1 x_{1}(t-\tau)}{(1+x_{1}(t-\tau)^c)},  \\
\dot{x}_i(t)&=&-\beta x_i(t)+\frac{\alpha_i x_{i}(t-\tau)}{(1+x_{i}(t-\tau)^c)} + \nonumber \\
& & C(x_{i-1}(t)-x_{i}(t)), \qquad i = 2, 3,\cdots, N,
\end{eqnarray}
\label{eq_1}
\end{subequations}
\begin{figure}
\centering
\includegraphics[width=1.0\columnwidth]{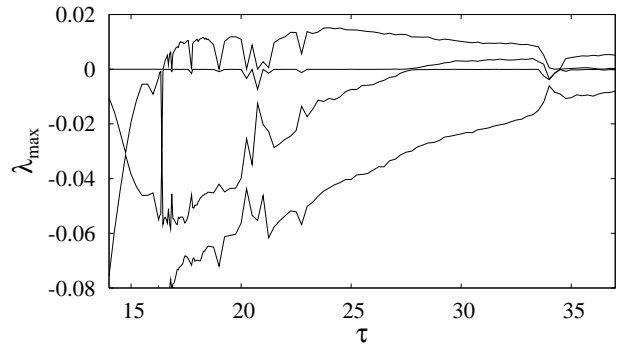}
\caption{\label{fig1b} The first four maximal Lyapunov exponents $\lambda_{max}$
of  the Mackey-Glass time-delay system (\ref{eq_1}a) for the parameter
values $\alpha=0.1, \beta=0.2, \tau \in(14,37)$ in the absence of the coupling $C$.}
\end{figure}
% 
%Here we apply the open end boundary conditions $x_0 = x_1$ 
%and $x_N = x_{N+1}$. 
where, $\alpha, \beta, c$ are the system parameters, $\tau$ is 
the time-delay and $C$ is the coupling strength. We have fixed the parameter 
values at $\alpha_1=0.2, \beta = 0.1$, $c = 10.0$, 
$\tau = 20.0$ and the values of the nonlinear parameter $\alpha_i$ of the 
response systems in the array are chosen randomly in the range  
$\alpha_{i} \in (0.17,0.20)$, so that all the systems are effectively 
nonidentical. For our simulations, we have fixed the total number of 
oscillators in the array as $N=20$, though we confirmed our results for $N = 50$
also (see Appendix A).

\begin{figure*}
\centering
\includegraphics[width=2.0\columnwidth]{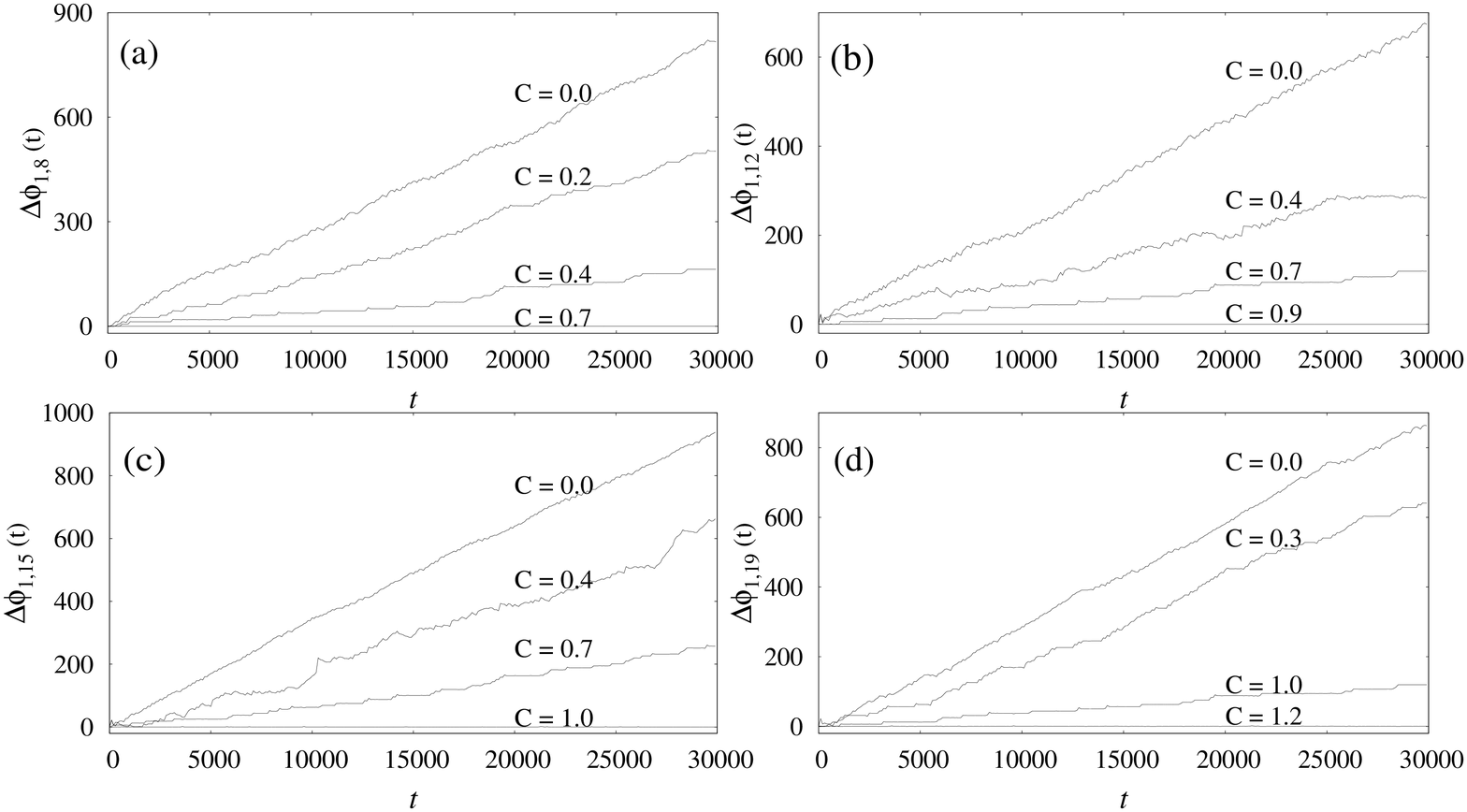}
\caption{\label{fig2} (a-d) Phase differences ($\Delta\phi_{1,i}$ = $\phi_{1}-\phi_{i}$) 
of the randomly selected systems ($i$ = 8,12,15,19) from the array of
coupled Mackey-Glass time-delay systems
for different values of the coupling strength $C$. A more detailed description can be found in
the text.}
\end{figure*}

The uncoupled drive (\ref{eq_1}a) exhibits a highly non-phase-coherent chaotic 
attractor for the chosen parameter values, which is depicted in Fig.~\ref{fig1}a. 
The first four largest Lyapunov exponents of the uncoupled drive system are shown in 
Fig.~\ref{fig1b} as a function of the delay time $\tau$. Note that
the phase calculated directly from the original non-phase-coherent chaotic attractor
(Fig.~\ref{fig1}a) cannot yield monotonically increasing behaviour as it has several closed
loops, which also contribute to the phase information, other than the main 
center of rotation of the major part of the trajectories~\cite{dvskml2006,dvskml2008}.
To overcome this problem a nonlinear transformation is introduced so as to rescale the original 
non-phase-coherent chaotic  attractor into smeared limit cycle-like attractor
with a single center of rotation. The transformation is performed by introducing
a new state variable~\cite{dvskml2006,dvskml2008},
\begin{equation}
z(t+\tau) = x(t)x(t+\hat{\tau})/x(t+\tau),
\label{eq_2}
\end{equation}
where $\hat{\tau}$ is the optimal value of time-delay to be chosen in order
to avoid any additional center of rotation. This functional form 
of transformation (along with a delay time $\hat\tau$) has been identified by 
generalizing the transformation used in the case of chaotic attractors in the 
Lorenz systems~\cite{aspmgr2001}. Now, the projected trajectory in
the new state space ($x(t+\tau),z(t+\tau)$) (Fig.~\ref{fig1}b) resembles 
that of a smeared limit cycle-like attractor with a single fixed center of rotation.
It is also to be noted that, even though the transformed attractor has sharp
turns in the vicinity of the common center, it does not have any closed loops
as in the original non-phase-coherent attractor. Otherwise, the transformed attractor
would not give rise to monotonically increasing phase resulting in exact matching
of the phases of the coupled time-delay systems~\cite{dvskml2006,dvskml2008}. 

\section{\label{sec:pc}GPS from the transformed attractor}
In this section, we will show that the global phase synchronization in the array 
of Mackey-Glass time-delay systems (\ref{eq_1}) is attained by a
sequential phase synchronization of the oscillators in the array as the coupling
strength is increased.  Further we will also demonstrate that the remaining 
non-synchronized oscillators in the array form synchronized clusters among themselves
before attaining GPS. 

We use the same nonlinear transformation (\ref{eq_2}) to recast the original non-phase-coherent
chaotic attractors of all the $N=20$ oscillators into smeared limit cycle-like attractors.
We have fixed the value of the optimal value $\hat{\tau}$ in Eq.~(\ref{eq_2}) as $\hat{\tau}=8.0$ for all the $N$ oscillators.
Instead one can also choose different values for $\hat{\tau}$ for different 
oscillators, as they are nonidentical systems with a parameter mismatch, to obtain
more exact unfolding for different attractors. However, we find $\hat\tau$ = 8.0 for 
all the oscillators is adequate for our purpose in the following study. We have calculated the instantaneous phases of
all the oscillators using the Poincar\'e section technique~\cite{aspmgr2001,sbjk2002} 
from their corresponding transformed attractors. Projected trajectories of
randomly selected response systems ($i$ = 8,12,15,19) in the array (\ref{eq_1}b) into 
the new state space ($x_i(t+\tau),z_i(t+\tau)$), where they look like smeared limit 
cycle-like attractors with a fixed center of rotation, are shown in Figs.~\ref{fig1}(c-f).
Filled circles in these plots correspond to the Poincar\'e points. 

\begin{figure}
\centering
\includegraphics[width=1.0\columnwidth]{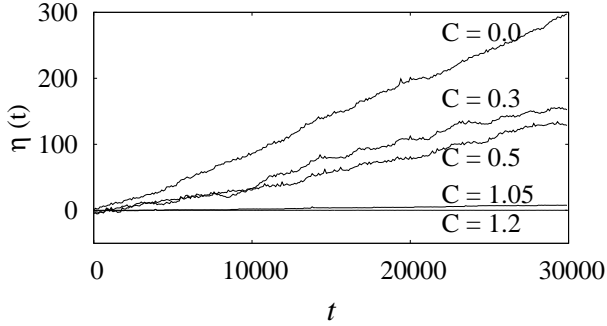}
\caption{\label{fig2a} The average phase difference ($\eta(t)$)
for different values of the coupling strength $C$. For $C=0.0$ the phases of all the 
systems are unbounded so the phase difference increases linearly with time but for $C=1.2$
the phases of all the systems are bounded, showing a high quality phase synchronization.}
\end{figure}

Phase differences, $\Delta\phi_{i} = \phi_{1}-\phi_{i}$, between the drive and some randomly selected response systems 
($i = 8,12,15,19$ ) in the array (\ref{eq_1}b) are shown in Figs.~\ref{fig2}
for different values of the coupling strength. They increase monotonically in the
absence of coupling ($C$ = 0.0) indicating that all the oscillators are in an asynchronous state. 
Phase slips in the phase differences for small values of the coupling strength 
indicates that the oscillators are in the transition state to GPS. 
Further increase in the value of the coupling strength results in a
strong boundedness of the phases of the oscillators. For sufficiently
large $C$, the phase differences become zero 
(Figs.~\ref{fig2}) indicating the existence of phase synchronization 
between the drive and the response systems. It is evident from the 
Figs.~\ref{fig2} that the $8$th oscillator is synchronized with the drive at 
$C = 0.7$, while the other systems are in the transition state, whereas
$12$th oscillator is synchronized with drive only at $C = 0.9$. The other
two oscillators with the index $i=15$  and $i=19$ reach synchronization
with the drive for further larger values of the coupling strength, $C = 1.0$ 
and $1.2$, respectively. 
Therefore, it is clear that the nearby oscillators to the drive system
in the array are synchronized first as the coupling strength is
increased implying that the global phase synchronization is reached
by sequential phase synchronization of the coupled oscillators 
in the array. 
\begin{figure}
\centering
\includegraphics[width=1.0\columnwidth]{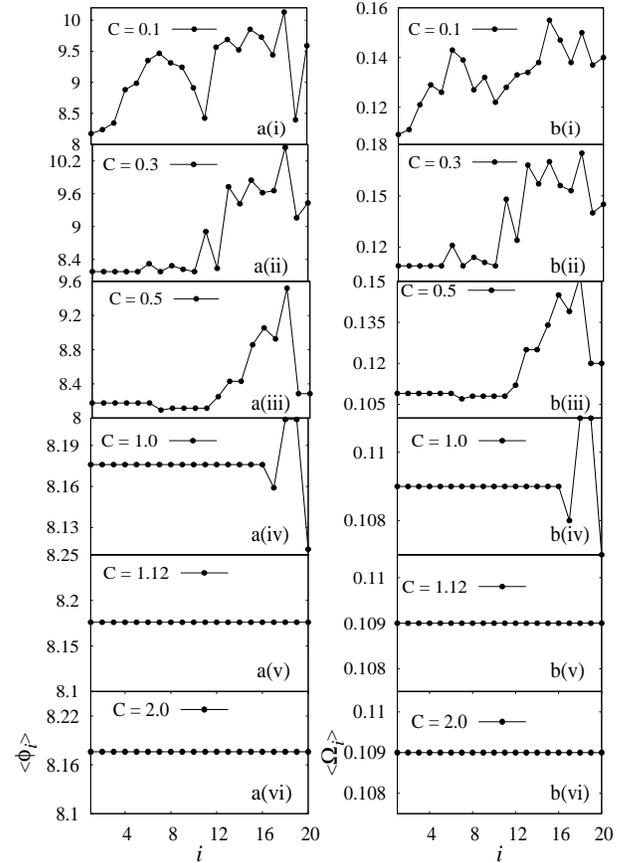}
\caption{\label{fig3} (a) Time averaged phase ($\langle \phi_i \rangle$) and (b) time averaged 
frequency ($\langle \Omega_i \rangle$) of all the systems plotted as a function of 
the system index $i$ for various values of the coupling strength $C$.}
\end{figure}
To confirm the existence of GPS, we have calculated the average phase
difference, $\eta(t)$, defined as
\begin{equation}
\eta(t) = \frac{1}{N-1} \sum_{j=2}^{N} (\phi_{1}-\phi_{j}).
\label{eq_3}
\end{equation}
The average phase difference ($\eta(t)$) for different values $C$ is shown in Fig.~\ref{fig2a} as a function of time $t$. 
In the absence of the coupling ($C=0.0$), the phases of all the oscillators evolve 
independently and hence their average phase difference increases linearly with time. 
Further increase in $C$ induces the entrainment of phases of 
the oscillators and at the value of coupling strength $C=1.2$ the average phase difference 
of all the $N$ oscillators becomes exactly zero, showing a high quality GPS in the 
array.
\begin{figure*}
\centering
\includegraphics[width=1.5\columnwidth]{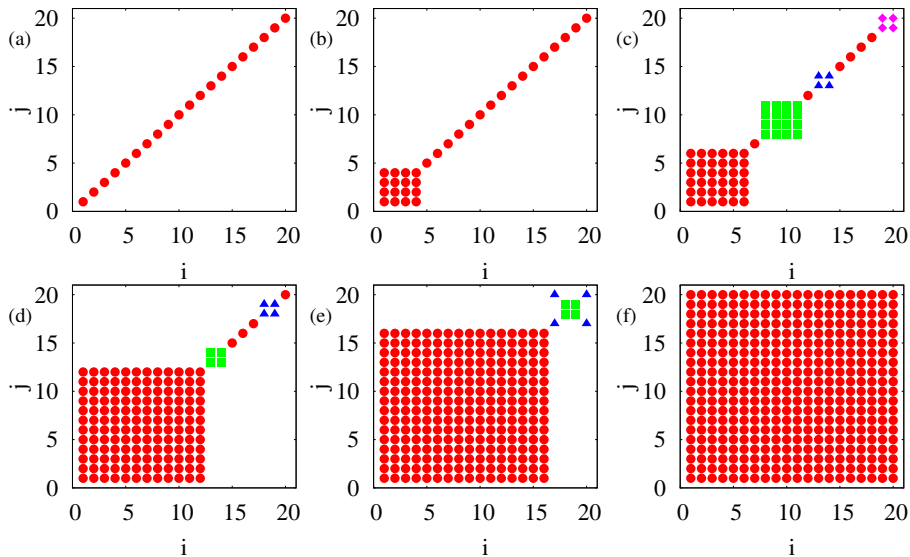}
\caption{\label{fig5} (Colour online) Snap shots of the node vs node diagrams 
(that is oscillator index vs oscillator index plots) indicating the
sequential phase synchronization and the
organization of cluster states for different values of coupling strength. 
The different symbols indicate that the corresponding nodes are phase 
synchronized. (a) non phase synchronized case for $C$ = 0.1 (b) First four oscillators 
in Eq. (\ref{eq_1}b) are phase synchronized with the drive system for $C$ = 0.27. (c), 
(d) and (e) Sequential phase synchronization and the formation 
of small cluster states for $C$ = 0.5, 0.82 and 1.08, respectively, and 
(f) Global phase synchronization for $C$ = 1.2.}
\end{figure*}

The emergence of GPS through a sequential phase synchronization is also
characterized by calculating the time averaged phase ($\langle \phi_i \rangle$) and the
time averaged frequency ($\langle \Omega_i \rangle$) of each of the oscillators in the
array which are defined as
\begin{subequations}
\begin{eqnarray}
\big\langle \phi_{i}(t)\big\rangle&=&\Big\langle2\pi k+2\pi \frac{t^{i}-t^{i}_{k}}{t^{i}_{k+1}-t^{i}_{k}}\Big\rangle_{t},~(t^{i}_{k}<t^{i}<t^{i}_{k+1})\\
\big\langle \Omega_{i}(t)\big\rangle &=& \lim_{T \to \infty} \frac{1}{T} \int_{0}^{T} \dot\phi_{i}(t) dt,
\end{eqnarray}
\label{eq3a}
\end{subequations}
where $t^{i}_{k}$ is the time of the $k^{th}$ crossing of the flow with the  Poincar\'e 
section of the $i^{th}$ attractor and $\langle...\rangle_{t}$ denotes a time average. 
The average phase and the average frequency are shown in Figs.~\ref{fig3}a
and Figs.~\ref{fig3}b, respectively, for different values of the coupling
strength as a function of the oscillator index. A random distribution of the
average phase (Fig.~\ref{fig3}a(i)) and the average frequency 
(Fig.~\ref{fig3}b(i)) for the value of the coupling strength
$C=0.1$ indicate that the coupled oscillators in the array evolve almost independently.
A slight increase in the coupling strength (to $C=0.3$) results in synchronous evolution of
the first $5$ oscillators in the array as seen in Fig.~\ref{fig3}a(ii) 
and Fig.~\ref{fig3}b(ii). For $C=0.5$, 
 Fig.~\ref{fig3}a(iii) and Fig.~\ref{fig3}b(iii) indicate that the first $6$ oscillators
are synchronized. It is also to be noted from these figures that the other 
desynchronized oscillators form synchronized clusters among themselves. 
In particular, the oscillators with the indices $8-11$ synchronize among themselves 
to form a cluster, while the oscillators with the indices $13-14$, and $19-20$ form separate small clusters.
These clusters can also be clearly visualized by plotting the oscillator index plots as
we will illustrate below. It is also to be noted that even when the total number of
oscillators in the array is increased, the phenomenon remains qualitatively the
same though the sizes of the clusters will increase appropriately
 (see Appendix A below for some details
for $N$ = 50).  We also note that the results remain qualitatively unaltered 
even for different sets of random values
for the nonlinear parameters, $\alpha_i$, confirming the robustness of our results. 
A similar transition to PS through clustering, termed as hard 
transition for large coupling strength, in a chain of diffusively coupled R\"ossler 
oscillators with large frequency mismatch have been observed~\cite{gvoasp1997} but in
the periodic state due to the suppression of chaotic attractors. For further larger values of
the coupling strength, the desynchronized oscillators form similar small clusters
among themselves before attaining GPS. The average phase and the average frequency
illustrated in  Fig.~\ref{fig3}a(iv) and Fig.~\ref{fig3}b(iv) for $C=1.0$ indicates that most of the nearest oscillators are
synchronized with the drive, while the oscillators with the indices $18-19$ form 
a small separate cluster. All the oscillators in the array become phase/frequency locked and evolve
in synchrony (GPS) with each other for the coupling strength $C=1.12$ as depicted in 
Fig.~\ref{fig3}a(v) and Fig.~\ref{fig3}b(v) and they continue to be in a stable 
phase/frequency synchronized state which is shown in Fig.~\ref{fig3}a(vi) and 
Fig.~\ref{fig3}b(vi) for $C = 2.0$. 

The mechanism for the formation of clusters and GPS may be explained
as follows. Due to the mismatches in the nonlinear parameters, $\alpha_i$,
all the individual systems  in the array evolve
independently with different phases (phase mismatches), and correspondingly with frequency
mismatches, for small values of the coupling strength $C$.  As $C$ is increased further, the
oscillators with nearest frequencies in the array synchronize first to form clusters
among themselves leaving the clusters with relatively large frequency mismatch in 
isolation (see Figs.~\ref{fig5} and~\ref{fig14}).
Further increase in $C$ results in the formation of a single large
cluster whose constituents exhibit a coherent phase oscillation with the drive due to
the decomposition of the clusters away from the drive in the array. 
Consequently GPS results in the system. Similar mechanism has been
identified in ensemble of coupled R\"ossler oscillators with frequency
mismatches~\cite{gvocsz2007} (without time-delay).

The above dynamical organization of GPS via sequential phase synchronization
and the clustering can be also be visualized clearly by using snap shots of 
the oscillators in the index vs index plot, as node vs node diagrams, 
as shown in  Fig.~\ref{fig5}. The oscillators that evolve with 
identical values of the average phase/frequency are assigned with identical 
shapes. The diagonal line in Fig.~\ref{fig5}a for $C=0.1$
corresponds to the oscillator index $i=j$ and they evolve independently.
Figure~\ref{fig5}b indicates that the first four oscillators in the array 
are synchronized with the drive for $C=0.27$. As discussed above, three 
small clusters are seen in Fig.~\ref{fig5}c for $C=0.5$ while the first $6$ 
oscillators form a large synchronized cluster.
Similar small clusters are shown in Figs.~\ref{fig5}d and \ref{fig5}e 
for $C=0.82$ and $1.08$, respectively, in addition to the single
large cluster formed by sequential phase synchronization of the oscillators in 
the array. Finally, GPS of all the oscillators in the array is illustrated 
in Fig.~\ref{fig5}f for $C=1.2$.

\begin{figure}
\centering
\includegraphics[width=1.0\columnwidth]{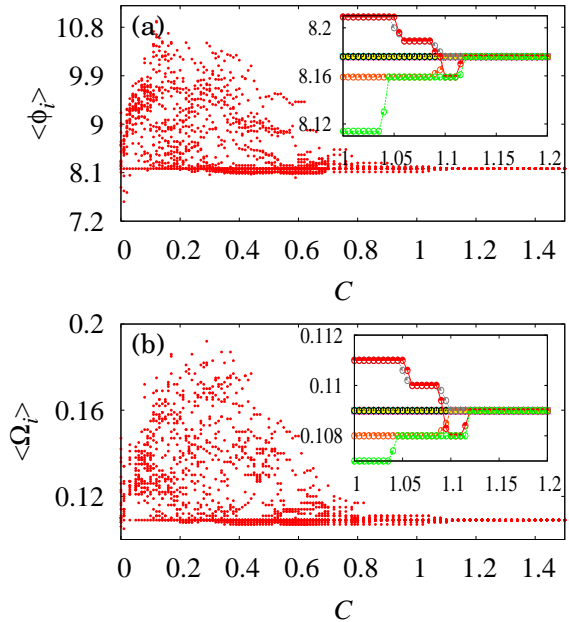}
\caption{\label{fig6} (Colour online) (a) Time averaged phase ($\langle \phi_i \rangle$) and 
(b) Time averaged frequency ($\langle \Omega_i \rangle$), $i = 1,2,\cdots, 20$  plotted as a function of 
the coupling strength $C \in (0,1.5)$. Here for each value of $C$
we have plotted the average phase/frequency of all the $N$ = 20 oscillators which is shown 
by the filled circles. Insets show some of the systems get synchronized 
themselves to form small clusters (each subgroup of oscillator is differentiated by
different types of circles) before they 
synchronize with the drive system to form GPS.}
\end{figure}

For a global picture of the emergence of GPS, we have plotted the 
average phase ($\langle \phi_i \rangle$)  and the average frequency 
($\langle \Omega_i \rangle$) of all the $N$ oscillators as a function of 
the coupling strength $C$ in Figs.~\ref{fig6}. There is an absence 
of any correlation among the average phases (Fig.~\ref{fig6}a) and 
the average frequencies (Fig.~\ref{fig6}b) of different oscillators for low values
of the coupling strength  as revealed by the random distributions of their values.
The random distribution of
$\langle \phi_i \rangle$ and $\langle \Omega_i \rangle$ are organized 
into few clusters for $C\ge 0.5$ as may be evident from the
Figs.~\ref{fig6}. Global phase synchronization emerges for 
$C\ge 1.12$ as may be seen from the insets. Small synchronized clusters formed by the 
remaining asynchronous oscillators for larger values of $C$ can also be
appreciated from the insets.

\begin{figure}
\centering
\includegraphics[width=1.0\columnwidth]{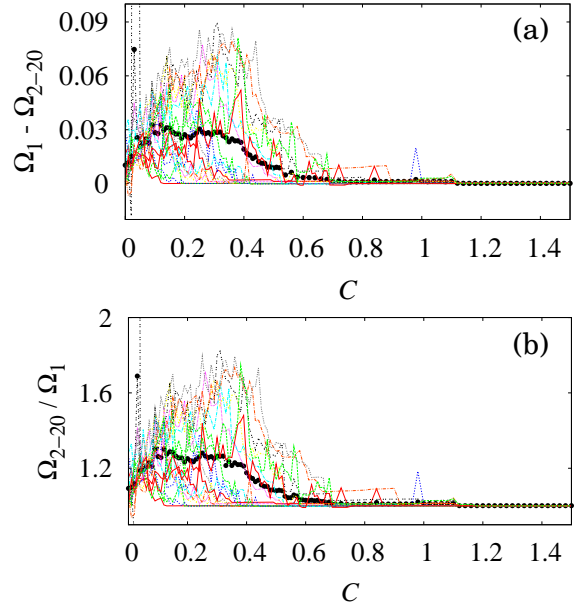}
\caption{\label{fig8} (Colour online) (a) The frequency difference ($\Delta \Omega_{1,j}, 
\quad j=2,3,\cdots,N$) and (b) the frequency ratio ($\Omega_j/\Omega_1, 
\quad j=2,3,\cdots,N$) are plotted as a function of the coupling strength 
$C \in (0,1.5)$. Each line corresponds to the difference/ratio between a 
response system and the drive system. The black filled circles indicate the average
frequency difference/ratio of all the ($N-1$) response systems from the drive system.}
\end{figure}

We have also plotted the  frequency difference ($\Delta \Omega_{1,j}, \quad j=2,3,\cdots,N$) 
and the frequency ratio ($\Omega_j/\Omega_1, \quad j=2,3,\cdots,N$) of all the oscillators 
with that of the drive as a function of the
coupling strength by different types of lines in Fig.~\ref{fig8}a and Fig.~\ref{fig8}b, respectively. 
The black filled circles in Fig.~\ref{fig8}a and Fig.~\ref{fig8}b correspond to the
average frequency difference and the average frequency ratio of all the oscillators with that
of the drive. The substantial saturation in their values for 
$C\ge 1.12$ indicates the emergence of GPS.

\begin{figure}
\centering
\includegraphics[width=1.0\columnwidth]{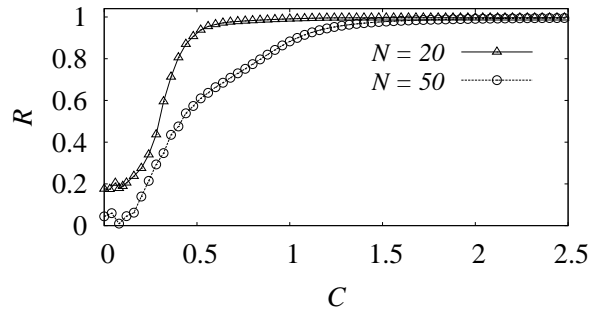}
\caption{\label{fig10} The phase order parameter ($R$) for the number of oscillators 
$N$ = 20 and $N$ = 50 as a function of 
the coupling strength indicating global phase synchronization in the array 
of coupled Mackey-Glass time-delay systems.}
\end{figure}

The emergence of global phase synchronization in the array can also be quantified
using the well-known order parameter~\cite{ymafp2004},
\begin{equation}
R~e^{i\psi} = \Big\langle\Big|\frac{1}{N}\sum_{j=1}^{N} e^{i\phi_{j}(t)}\Big|\Big\rangle_t
\label{eq_4}
\end{equation}
where $\phi_{j}(t)$ denotes the instantaneous phase of the $j^{th}$ system, $\psi(t)$ is the 
average phase and $\langle...\rangle_t$ denotes a time average. If all the systems are in a phase 
synchronized state then $R\approx 1$. The order parameter ($R$)
is plotted in Fig.~\ref{fig10} for the number of oscillators $N$ = 20 and $N$ = 50 
as a function of the coupling strength $C$. As $C$
is increased, $R$ also increases and for $N$ = 20 the critical value of the 
coupling is $C>1.12$ and for $N$ = 50, $C$ will be $> 2.4$, for the value of
$R\approx 1$ confirms the existence of GPS in the array of coupled time-delay systems.

\section{\label{sec:npc}GPS from the original non-phase-coherent attractor}
In this section, we demonstrate the existence of GPS from the original
non-phase-coherent chaotic attractors using two different approaches, namely 
recurrence quantification analysis~\cite{mcrmt2005,nmmcr2007} and the concept of localized 
sets~\cite{tpmsb2007} without estimating explicitly the measure of phase. 

\subsection{GPS using recurrence analysis}
Several measures of complexity which quantify small scale structures in the recurrence plots
have been proposed and are known as recurrence quantification analysis (RQA)~\cite{mcrmt2005,nmmcr2007}.
Certain measures have also been introduced to characterize and identify different kinds of
synchronization transitions in coupled chaotic systems. These measures have the advantage of 
applicability in the analysis of experimental systems and, in particular, in the case of very small 
available data sets.  Further, these measures can also be used in the case of  
non-phase-coherent chaotic/hyperchaotic attractors of time-delay 
systems~\cite{dvskml2006,dvskml2008,dvskmlejp2008}, where it is difficult and often even impossible
to calculate the phase explicitly. Among the available recurrence
quantification measures, we use  the Correlation of Probability of 
Recurrence (CPR) and the generalized autocorrelation function $P(t)$
to confirm the existence of GPS in the array of coupled time-delay systems (\ref{eq_1}),
both qualitatively and quantitatively.

A criterion to quantify phase synchronization between two systems
is the Correlation of Probability of Recurrence (CPR) defined as
\begin{align}
CPR=\langle \bar{P_1}(t)\bar{P_2}(t)\rangle/\sigma_1\sigma_2,
\end{align}
where $P(t)$ is the generalized autocorrelation function represented as
\begin{equation}
P(t)=\frac{1}{N-t} \sum_{i=1}^{N-t} \Theta(\epsilon-||X_i-X_{i+t}|| ),
\end{equation}
where $\Theta$ is the Heaviside function, $X_i$ is the $i^{th}$ data point of
the system $X$, $\epsilon$ is a predefined threshold, $|| . ||$ is the Euclidean 
norm, and $N$ is the number of data points, $\bar{P}_{1,2}$ means that the 
mean value has been subtracted and
$\sigma_{1,2}$ are the standard deviations of $P_1(t)$ and $P_2(t)$,
respectively.  Looking at the coincidence of the positions of the maxima of 
$P(t)$ of the systems, one can qualitatively identify PS~\cite{mcrmt2005,nmmcr2007}. 
If both  systems are in CPS, the probability of recurrence is
maximal at the same time $t$ and CPR $\approx 1$. If they are not in CPS,
the maxima do not occur simultaneously and hence one can expect a drift in both
the probability of recurrences resulting in low values of  CPR. 

\begin{figure*}
\centering
\includegraphics[width=1.7\columnwidth]{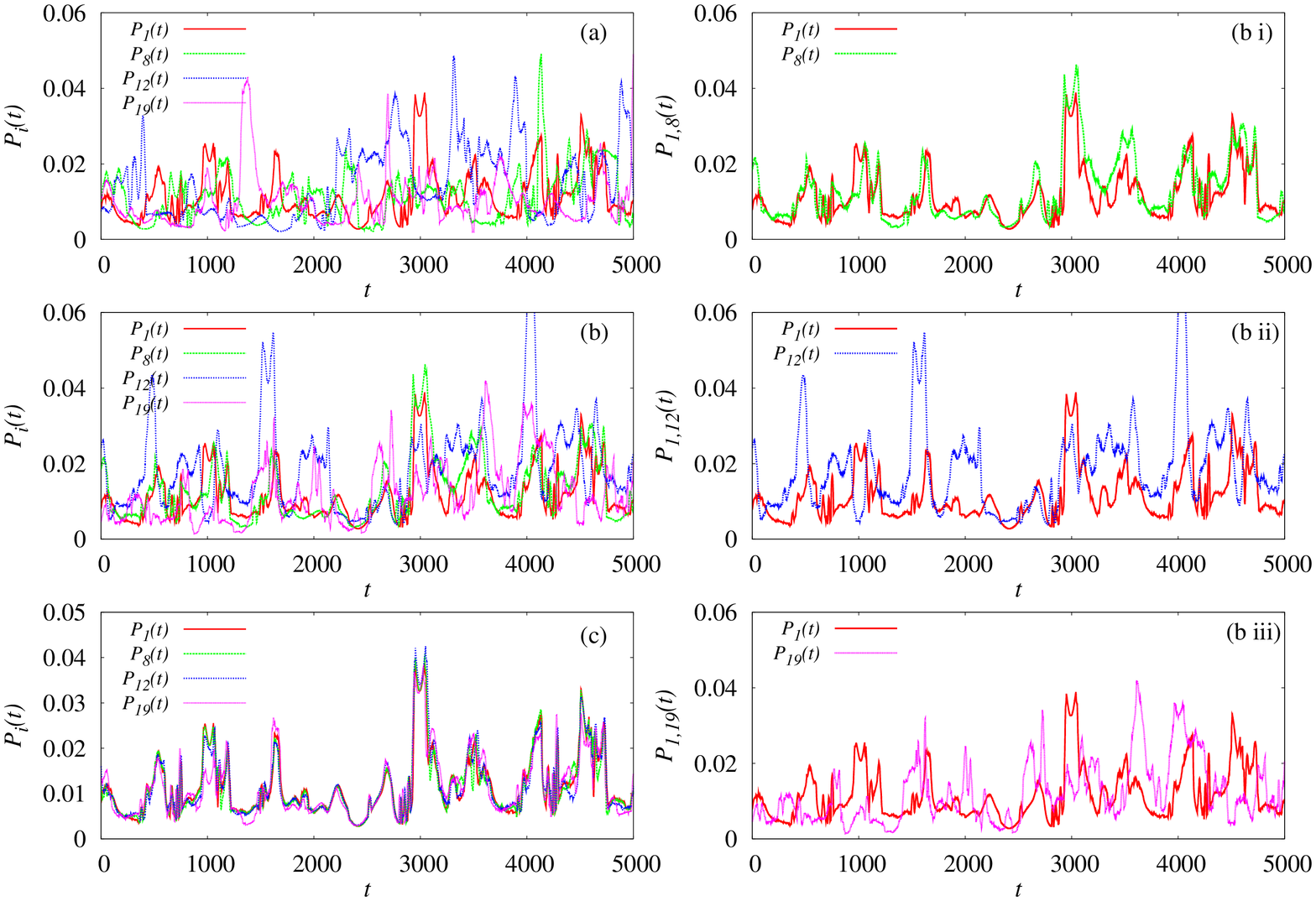}
\caption{\label{fig11} (Colour online) Generalized autocorrelation functions of the 
drive $P_{1}(t)$ and randomly selected response systems ($i=8,12$,and,$19$) $P_{8}(t)$, 
$P_{12}(t)$, and $P_{19}(t)$ indicating (a) Non-phase-synchronization for $C=0.0$,
(b) generalized autocorrelation functions  for $C=0.4$
(bi) PS  between the systems $1$ and $8$, (bii) approximate PS between the systems $1$ and
$12$ and (biii) non PS between the systems $1$ and $19$, and (c) PS between all 
the systems ($i=1,8,12$, and $19$) for $C=1.2$.}
\end{figure*}

The generalized autocorrelation function of the drive $P_1(t)$ and that of
some response systems ($i=8,12$, and $19$) $P_{8}(t)$, $P_{12}(t)$, and $P_{19}(t)$
are depicted in  Fig.~\ref{fig11} for different values of the coupling strength.
In the absence of coupling ($C=0.0$), all systems evolve independently and
hence the maxima of their respective generalized autocorrelation functions do not occur
simultaneously as shown in Fig.~\ref{fig11}a. On increasing the coupling strength,
oscillators with a lower value of index in the array become synchronized first resulting in 
sequential phase synchronization and this can also be identified from the
generalized autocorrelation functions of the response systems in the array.
For instance, $P_{8}(t)$, $P_{12}(t)$, and $P_{19}(t)$ are shown along with $P_{1}(t)$ in 
Fig.~\ref{fig11}b for $C=0.4$. It is clear
from this figure that the maxima of the drive $P_{1}(t)$ and those of the
response $P_{8}(t)$ are in complete agreement with each other (Fig.~\ref{fig11}bi) indicating the
existence of PS between them. On the other hand, only some of the maxima of the
response system $P_{12}(t)$ are in coincidence with those of the drive (Fig.~\ref{fig11}bii) illustrating
that the response system $i=12$ is in transition to PS,
whereas the maxima of the response system $P_{19}(t)$ do not coincide with
those of the drive (Fig.~\ref{fig11}biii) indicating that the response system $i=19$ is in an asynchronous 
state for the same value of $C$. For $C=1.2$, almost all of the positions of 
the peaks of the generalized auto correlation functions $P_{1}(t)$, $P_{8}(t)$, 
$P_{12}(t)$, and $P_{19}(t)$ are in agreement with each other as illustrated 
in Fig.~\ref{fig11}(c) confirming the existence of GPS via sequential phase
synchronization. It is also to be noted that the magnitudes 
of the peaks of all the oscillators have generally of different values and 
the differences in the heights of the peaks indicate that there is no 
correlation in the amplitudes of the coupled systems. Furthermore, the formation of
clusters by the other asynchronous oscillators in the array can also be realized by
plotting their respective generalized autocorrelation functions, which will show that
all their maxima are in good agreement with each other, whereas there exists a drift 
between them and the maxima of the sequentially synchronized cluster.

\begin{figure}
\centering
\includegraphics[width=1.0\columnwidth]{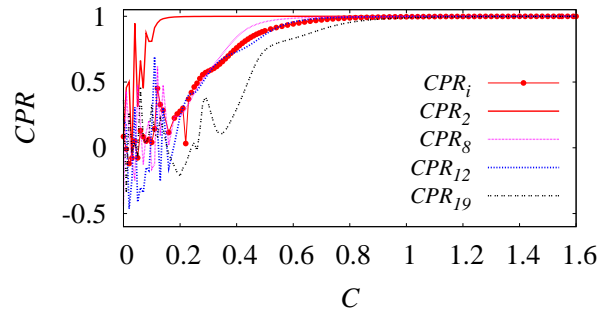}
\caption{\label{fig12} (Colour online) The index CPR as a function of the coupling strength $C$. 
Different lines correspond to the CPR of different ($i=2,8,12,$ and $19$) response 
systems with the drive system. 
The filled circles correspond to the mean value of the CPR of all the ($N-1$) systems 
in the array.}
\end{figure}

The existence of GPS via sequential phase synchronization is  also quantified 
using value of the index CPR of the response systems with the drive as shown in Fig.~\ref{fig12}. 
The different lines correspond to the index of the oscillators ($i$ = 2,8,12,19) in the
array. 
It is evident from the figure that the oscillators with increasing index attain
the value of unity in a sequence as a function of the coupling strength and finally
for $C>1.12$ the CPR of all the response systems with the drive reaches
unity confirming that all the coupled oscillators are in GPS.  
The mean value of CPR of all the response systems in the array is shown as filled circles, which
also confirms the existence of GPS for $C>1.12$.
\begin{figure*}
\centering
\includegraphics[width=2.5\columnwidth]{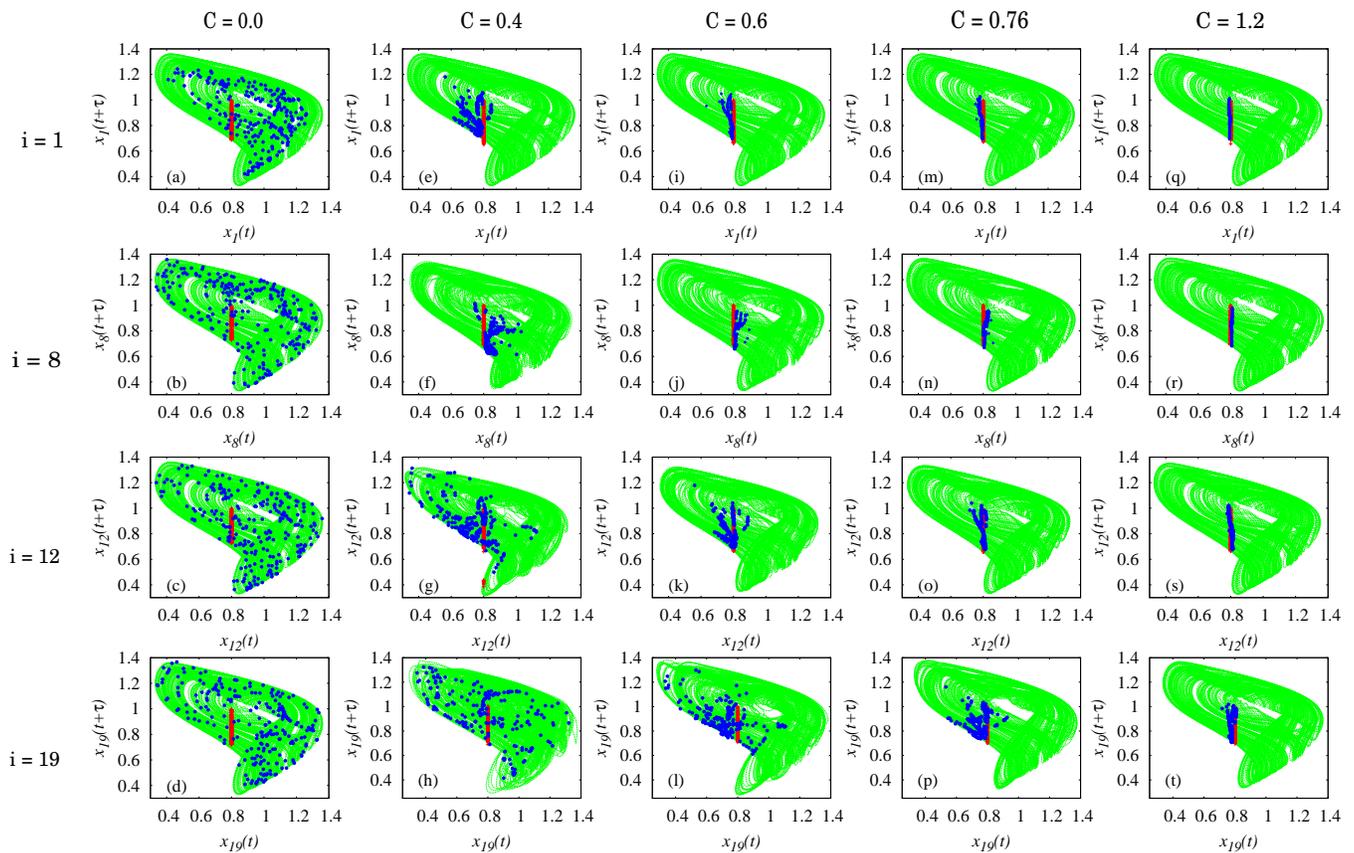}
\caption{\label{fig13} (Colour online) 
First row (a-q) corresponds to the attractors of the drive system ($i = 1$) 
and the rows (b-r, c-s, and d-t) correspond to the
attractors of some randomly selected response systems ($i = 8,12,19$). 
The $`+`$ marks represent the events (Poincar\'e sections) in the 
corresponding attractors. In (a-d) the sets (represented by the filled circles) 
are spread over the attractors and hence 
there is no CPS for the value of coupling strength $C$ = 0.0. In (e-h) for 
$C$ = 0.4 and in (i-l, m-p, q-t) the sets 
are localized confirming the existence of GPS in the array for $C$ = 0.6, 0.76, and 1.2, respectively.}
\end{figure*}

\subsection{GPS using the concept of localized sets}
Recently, an interesting framework to identify CPS, namely, the
concept of localized sets~\cite{tpmsb2007} has been introduced.
This approach provides an easy and efficient way to detect CPS especially
in complicated non-phase-coherent attractors. The basic
idea of this concept is to define a typical event in one of the 
systems and then observe the other system whenever this event occurs. 
These observations give raise to a set $D$. Depending upon the property 
of this set $D$, one can state whether PS exists or not. 
The coupled systems evolve independently if the sets obtained by 
observing the corresponding events in the systems spread over the 
attractor of the systems. On the other hand, if the sets are localized 
on the attractors then CPS exists between them.

We have confirmed the existence of the GPS in the linear array of 
Mackey-Glass time-delay systems (\ref{eq_1}) by using this concept of localized sets.
Now, we will demonstrate the existence of GPS via sequential phase synchronization
in the randomly selected response systems ($i$ = 1,8,12,19). We have defined
the event as Poinca\'re sections in the attractors indicated as $`+`$ marks
in Figs.~\ref{fig13}. 
The set, indicated as filled circles, obtained by observing the drive system 
($i = 1$) whenever the defined 
event occurs in the response system ($i = 8$) is shown in Fig.~\ref{fig13} (a) 
and that obtained by observing the response systems $i = 8,12,19$ whenever 
the defined event occurs in the drive system are shown in Figs.~\ref{fig13} (b-d) 
for the value of  coupling strength $C$ = 0.0. As the obtained sets are spread over
the attractors, all the systems evolve independently and there is no CPS in 
the absence of coupling between them. Further when we increase the coupling
strength to $C$ = 0.4, the oscillator ($i$ = 8) is partially synchronized with the drive
as the sets are almost localized but the sets in the oscillators $i$ = 12,19
are spread over the attractors which means that they are not yet phase synchronized 
with the drive system. This is shown in Figs.\ref{fig13} (e-h). Again increasing the 
coupling strength to $C$ = 0.6, the sets are further bounded to a small region over the 
attractors which shows that the oscillators $i$ = 8,12 are synchronized with the drive, but 
the oscillator $i$ = 19 is less phase synchronized with the drive 
which is represented by the spread of the events over the attractor as shown in Figs.\ref{fig13} (i-l). 
Further, the Figs.\ref{fig13} (m-p) and Figs.\ref{fig13} (q-t) indicate the situation for 
$C$ = 0.76 and $C$ = 1.2, respectively, where all the oscillators 
are now phase synchronized with the drive as 
the sets are localized over the attractor confirming the existence of GPS in an array via 
sequential phase synchronization as the coupling strength is increased.

Further, the formation of clusters can also be realized using the concept of
localized sets by defining the event among the response systems that form the clusters
and observing the other response systems that are in the same cluster. In this case 
the obtained sets by observing the event in the drive will spread over the attractor
of the response systems, while the sets obtained by observing the event among
the response systems that form a cluster will be localized on their respective 
attractors.
\begin{figure*}
\centering
\includegraphics[width=2.0\columnwidth]{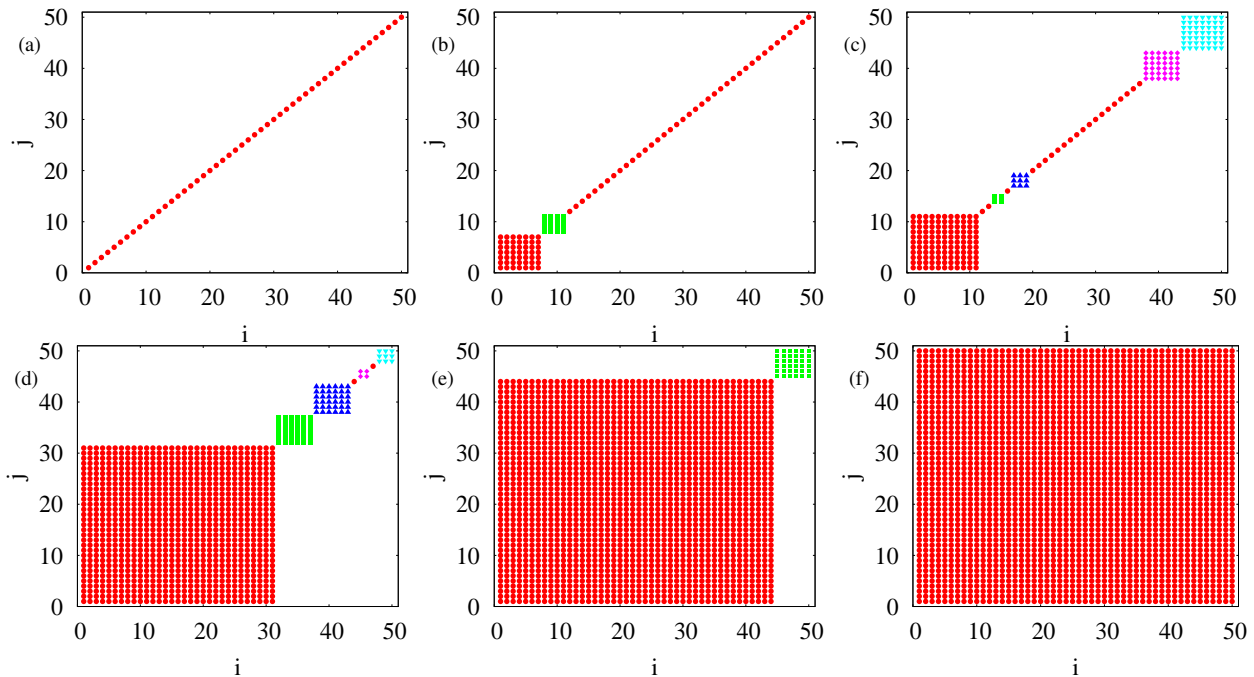}
\caption{\label{fig14} (Colour online) Snap shots of the node vs node diagram indicating the
sequential phase synchronization and the
organization of cluster states of $N=50$ oscillators for different values of coupling strength, $C$. 
The different symbols indicate that the corresponding nodes are phase 
synchronized. (a) Non-phase synchronized case for $C$ = 0.1, (b) First seven oscillators 
in Eq. (\ref{eq_1}b) are phase synchronized with the drive system for $C$ = 0.4, (c), 
(d) and (e) Sequential phase synchronization and the formation 
of small cluster states for $C$ = 0.53, 1.9 and 2.3, respectively, and 
(f)Global phase synchronization of $N=50$ oscillators for $C$ = 2.5.}
\end{figure*}
\section{\label{sec:con}Summary and Conclusion}

We have demonstrated the existence of global phase synchronization via
sequential phase synchronization in an array of coupled Mackey-Glass time-delay systems
with parameter mismatches, which exhibit highly non-phase-coherent attractors with
complex topological structure. Further, we have also shown that the remaining
asynchronous systems will organize
themselves to form different clusters before they get phase synchronized with the main
cluster to form global phase synchronization. We have confirmed the existence of
GPS via sequential phase synchronization by estimating the phase difference as a
function of the coupling strength, the average 
frequency and the average phase as a function of  the oscillator index and the
coupling strength after calculating the phase variables from the transformed 
attractors.  We have also demonstrated the existence of sequential phase synchronization
and the formation of clusters by the remaining oscillators using the
average frequency, average phase and specifically using the index vs index plot of the
oscillators.  Furthermore, we have also confirmed the existence of GPS
via sequential phase synchronization using  the recurrence quantification measures
and the concept of localized sets which are calculated from the original
non-phase-coherent attractors of the coupled Mackey-Glass time-delay systems.
It is also to noted that we have obtained similar
transitions to GPS via clustering even in the hyperchaotic regimes of Fig. 2 for the Mackey-Glass
systems and also in the coupled piecewise linear time-delay systems (with five positive 
Lyapunov exponents).
\begin{acknowledgments}
The work of R.S. and M.L. has been supported by the Department of Science 
and Technology (DST), Government of India sponsored IRHPA research project, 
and DST Ramanna program of M.L. D.V.S. has been supported by the Alexander von Humboldt 
Foundation. J.K. acknowledges the support from  EU under project No. 240763 PHOCUS(FP7-ICT-2009-C).
\end{acknowledgments}
\appendix
\section{\label{a1} }

The dynamical organization of GPS via sequential phase synchronization and the cluster 
formation can be visualized by using the snap shots of $N = 50$ oscillators in 
the index vs index plot as shown in Fig.\ref{fig14}. The diagonal line in Fig.\ref{fig14} (a)
for the value of coupling $C = 0.1$ corresponds to the oscillator index $i=j$ and the oscillators
evolve independently. Further in Fig.\ref{fig14} (b), the first seven oscillators in the 
array are synchronized with the drive and the oscillators $8-11$ form a separate 
cluster for the coupling strength $C=0.4$. Further, the first eleven oscillators in the 
array are synchronized and four small separate clusters 
are seen in Fig.\ref{fig14} (c) for $C=0.53$. Similar small 
clusters are formed in Fig.\ref{fig14} (d) for $C=1.9$ while the first thirty one
oscillators form a large synchronized cluster. Further, in Fig.\ref{fig14} (e) the first forty four oscillators 
are synchronized with the drive and the oscillators $45-50$ form a separate cluster 
for $C=2.3$. Finally, the occurrence of GPS of all the oscillators in the array is illustrated in 
Fig.\ref{fig14} (f) for the value of coupling strength $C=2.5$.


\begin{thebibliography}{26}
\expandafter\ifx\csname natexlab\endcsname\relax\def\natexlab#1{#1}\fi
\expandafter\ifx\csname bibnamefont\endcsname\relax
  \def\bibnamefont#1{#1}\fi
\expandafter\ifx\csname bibfnamefont\endcsname\relax
  \def\bibfnamefont#1{#1}\fi
\expandafter\ifx\csname citenamefont\endcsname\relax
  \def\citenamefont#1{#1}\fi
\expandafter\ifx\csname url\endcsname\relax
  \def\url#1{\texttt{#1}}\fi
\expandafter\ifx\csname urlprefix\endcsname\relax\def\urlprefix{URL }\fi
\providecommand{\bibinfo}[2]{#2}
\providecommand{\eprint}[2][]{\url{#2}}

\bibitem[{\citenamefont{Pikovsky et~al.}(2001)}]{aspmgr2001}
\bibinfo{author}{\bibfnamefont{A.~S.} \bibnamefont{Pikovsky}},
  \bibinfo{author}{\bibfnamefont{M.~G.} \bibnamefont{Rosenblum}},
  \bibnamefont{and} \bibinfo{author}{\bibfnamefont{J.}~\bibnamefont{Kurths}},
  \emph{\bibinfo{title}{Synchronization - A Unified Approach to Nonlinear
  Science}} (\bibinfo{publisher}{Cambridge University Press},
  \bibinfo{address}{Cambridge}, \bibinfo{year}{2001}).

\bibitem[{\citenamefont{Boccaletti et~al.}(2002)}]{sbjk2002}
\bibinfo{author}{\bibfnamefont{S.}~\bibnamefont{Boccaletti}},
  \bibinfo{author}{\bibfnamefont{J.}~\bibnamefont{Kurths}},
  \bibinfo{author}{\bibfnamefont{G.}~\bibnamefont{Osipov}},
  \bibinfo{author}{\bibfnamefont{D.~L.} \bibnamefont{Valladares}},
  \bibnamefont{and} \bibinfo{author}{\bibfnamefont{C.~S.} \bibnamefont{Zhou}},
  \bibinfo{journal}{Phys. Rep.} \textbf{\bibinfo{volume}{366}},
  \bibinfo{pages}{1} (\bibinfo{year}{2002}).

\bibitem[{\citenamefont{issue on Phase synchronization edited~by
  J.~Kurths}(2000)}]{jk2000}
\bibinfo{journal}{Int. J. Bifurcation Chaos Appl. Sci. Eng.}
  \textbf{\bibinfo{volume}{10}} (\bibinfo{year}{2000}),
\bibinfo{author}{\bibfnamefont{special}~\bibnamefont{issue on Phase synchronization
  edited~by J.~Kurths}}.

\bibitem[{\citenamefont{Ivanchenko et~al.}(2004)}]{mvigvo2004}
\bibinfo{author}{\bibfnamefont{M. V.} \bibnamefont{Ivanchenko}}, 
\bibinfo{author}{\bibfnamefont{G. V.} \bibnamefont{Osipov}}, 
\bibinfo{author}{\bibfnamefont{V. D.} \bibnamefont{Shalfeev}}, 
\bibnamefont{and}  \bibinfo{author}{\bibfnamefont{J.} \bibnamefont{Kurths}}, 
  \bibinfo{journal}{Phys. Rev. Lett.} \textbf{\bibinfo{volume}{93}},
  \bibinfo{pages}{134101} (\bibinfo{year}{2004}).

\bibitem[{\citenamefont{Takamatsu et~al.}(2000)}]{attf2000}
\bibinfo{author}{\bibfnamefont{A.} \bibnamefont{Takamatsu}}, 
  \bibinfo{author}{\bibfnamefont{T.} \bibnamefont{Fujii}}, 
\bibnamefont{and}  \bibinfo{author}{ \bibnamefont{I.} \bibnamefont{Endo}}, 
  \bibinfo{journal}{Phys. Rev. Lett} \textbf{\bibinfo{volume}{85}},
  \bibinfo{pages}{2026} (\bibinfo{year}{2000}).

\bibitem[{\citenamefont{Osipov et~al.}(1996)}]{aspmgr1996}
  \bibinfo{author}{\bibfnamefont{A. S} \bibnamefont{Pikovsky}},
  \bibinfo{author}{\bibfnamefont{M.~G.} \bibnamefont{Rosenblum}},
\bibnamefont{and}  \bibinfo{author}{\bibfnamefont{J.} \bibnamefont{Kurths}}, 
  \bibinfo{journal}{Europhys. Lett.} \textbf{\bibinfo{volume}{34}},
  \bibinfo{pages}{165} (\bibinfo{year}{1996}).

\bibitem[{\citenamefont{kiss et~al.}(2002)}]{izkyz2002}
\bibinfo{author}{\bibfnamefont{I. Z.} \bibnamefont{Kiss}}, 
\bibinfo{author}{\bibfnamefont{Y.} \bibnamefont{Zhai}}, 
\bibnamefont{and}  \bibinfo{author}{\bibfnamefont{J. L.} \bibnamefont{Hudson}}, 
  \bibinfo{journal}{Phys. Rev. Lett.} \textbf{\bibinfo{volume}{88}},
  \bibinfo{pages}{238301} (\bibinfo{year}{2002}).

\bibitem[{\citenamefont{kiss et~al.}(2002)}]{czjk2002}
\bibinfo{author}{\bibfnamefont{C.} \bibnamefont{Zhou}}, 
\bibinfo{author}{\bibfnamefont{J.} \bibnamefont{Kurths}}, 
\bibinfo{author}{\bibfnamefont{I. Z.} \bibnamefont{Kiss}}, 
\bibnamefont{and}  \bibinfo{author}{\bibfnamefont{J. L.} \bibnamefont{Hudson}}, 
  \bibinfo{journal}{Phys. Rev. Lett.} \textbf{\bibinfo{volume}{89}},
  \bibinfo{pages}{014101} (\bibinfo{year}{2002}).

\bibitem[{\citenamefont{Osipov et~al.}(1997)}]{gvoasp1997}
\bibinfo{author}{\bibfnamefont{G.~V.} \bibnamefont{Osipov}}, 
  \bibinfo{author}{\bibfnamefont{A.~S.} \bibnamefont{Pikovsky}}, 
  \bibinfo{author}{\bibfnamefont{M.~G.} \bibnamefont{Rosenblum}},
\bibnamefont{and}  \bibinfo{author}{\bibfnamefont{J.} \bibnamefont{Kurths}}, 
  \bibinfo{journal}{Phys. Rev. E.} \textbf{\bibinfo{volume}{55}},
  \bibinfo{pages}{2353} (\bibinfo{year}{1997}).

\bibitem[{\citenamefont{zheng et~al.}(1997)}]{zheng1998}
\bibinfo{author}{\bibfnamefont{Z.} \bibnamefont{Zheng}}, 
  \bibinfo{author}{\bibfnamefont{G.} \bibnamefont{Hu}}, 
\bibnamefont{and}  \bibinfo{author}{\bibfnamefont{B.} \bibnamefont{Hu}}, 
  \bibinfo{journal}{Phys. Rev. Lett.} \textbf{\bibinfo{volume}{81}},
  \bibinfo{pages}{5318} (\bibinfo{year}{1998}).

\bibitem[{\citenamefont{Zhan et~al.}(2000)}]{mzzz2000}
\bibinfo{author}{\bibfnamefont{M.} \bibnamefont{Zhan}}, 
  \bibinfo{author}{\bibfnamefont{Z.~G} \bibnamefont{Zheng}}, 
  \bibinfo{author}{\bibfnamefont{G.} \bibnamefont{Hu}}, 
\bibnamefont{and}  \bibinfo{author}{ \bibnamefont{Xi-hong Peng}}, 
  \bibinfo{journal}{Phys. Rev. E} \textbf{\bibinfo{volume}{62}},
  \bibinfo{pages}{3552} (\bibinfo{year}{2000}).

\bibitem[{\citenamefont{Kozyreff et~al.}(2000)}]{gkagv2000}
\bibinfo{author}{\bibfnamefont{G.} \bibnamefont{Kozyreff}}, 
\bibinfo{author}{\bibfnamefont{A.~G.} \bibnamefont{Vladimirov}}, 
\bibnamefont{and}  \bibinfo{author}{\bibfnamefont{P.} \bibnamefont{Mandel}}, 
  \bibinfo{journal}{Phys. Rev. Lett.} \textbf{\bibinfo{volume}{85}},
  \bibinfo{pages}{3809} (\bibinfo{year}{2000}).

\bibitem[{\citenamefont{Otsuka et~al.}(2000)}]{kork2000}
\bibinfo{author}{\bibfnamefont{K.} \bibnamefont{Otsuka}}, 
\bibinfo{author}{\bibfnamefont{Y.} \bibnamefont{Miyasaka}}, 
\bibinfo{author}{\bibfnamefont{T.} \bibnamefont{Narita}}, 
\bibinfo{author}{\bibfnamefont{S.~C.} \bibnamefont{Chu}}, 
\bibinfo{author}{\bibfnamefont{C.~C.} \bibnamefont{Lin}},
\bibnamefont{and}  \bibinfo{author}{\bibfnamefont{J.~Y.} \bibnamefont{Ko}}, 
  \bibinfo{journal}{Phys. Rev. Lett.} \textbf{\bibinfo{volume}{97}},
  \bibinfo{pages}{213901} (\bibinfo{year}{2006}).

\bibitem[{\citenamefont{Boccaletti et~al.}(2006)}]{sbvl2006}
\bibinfo{author}{\bibfnamefont{S.} \bibnamefont{Boccaletti}},
\bibinfo{author}{\bibfnamefont{V.} \bibnamefont{Latora}},
\bibinfo{author}{\bibfnamefont{Y.} \bibnamefont{Moreno}},
\bibinfo{author}{\bibfnamefont{M.} \bibnamefont{Chavez}},
  \bibnamefont{and}
  \bibinfo{author}{\bibfnamefont{D.~U.}~\bibnamefont{Hwang}},
  \bibinfo{journal}{Phys. Rep.} \textbf{\bibinfo{volume}{424}},
  \bibinfo{pages}{175} (\bibinfo{year}{2006}). 

\bibitem[{\citenamefont{Arenas et~al.}(2008)}]{aaadl2008}
\bibinfo{author}{\bibfnamefont{A.} \bibnamefont{Arenas}},
\bibinfo{author}{\bibfnamefont{A.} \bibnamefont{Díaz-Guilera}},
\bibinfo{author}{\bibfnamefont{J.} \bibnamefont{Kurths}},
\bibinfo{author}{\bibfnamefont{Y.} \bibnamefont{Moreno}},
  \bibnamefont{and}
  \bibinfo{author}{\bibfnamefont{C.}~\bibnamefont{Zhou}},
  \bibinfo{journal}{Phys. Rep.} \textbf{\bibinfo{volume}{469}},
  \bibinfo{pages}{93} (\bibinfo{year}{2008}). 

\bibitem[{\citenamefont{Batista et~al.}(2007)}]{casbamb2007}
\bibinfo{author}{\bibfnamefont{C. A. S.} \bibnamefont{Batista}},
\bibinfo{author}{\bibfnamefont{A. M.} \bibnamefont{Batista}},
\bibinfo{author}{\bibfnamefont{J. A. C.} \bibnamefont{dePontes}},
\bibinfo{author}{\bibfnamefont{R. L.} \bibnamefont{Viana}},
  \bibnamefont{and}
  \bibinfo{author}{\bibfnamefont{S. R.}~\bibnamefont{Lopes}},
  \bibinfo{journal}{Phys. Rev. E} \textbf{\bibinfo{volume}{76}},
  \bibinfo{pages}{016218} (\bibinfo{year}{2007}). 

\bibitem[{\citenamefont{Ren et~al.}(2007)}]{qrjz2007}
\bibinfo{author}{\bibfnamefont{Q.} \bibnamefont{Ren}}, 
\bibnamefont{and}  \bibinfo{author}{\bibfnamefont{J.} \bibnamefont{Zhao}}, 
  \bibinfo{journal}{Phys. Rev. E} \textbf{\bibinfo{volume}{76}},
  \bibinfo{pages}{016207} (\bibinfo{year}{2007});
\bibinfo{author}{\bibfnamefont{X.} \bibnamefont{Yu}}, 
\bibinfo{author}{\bibfnamefont{Q.} \bibnamefont{Ren}}, 
\bibinfo{author}{\bibfnamefont{J.} \bibnamefont{Hou}}, 
\bibnamefont{and}  \bibinfo{author}{\bibfnamefont{J.} \bibnamefont{Zhao}}, 
  \bibinfo{journal}{Phys. Lett. A} \textbf{\bibinfo{volume}{373}},
  \bibinfo{pages}{1276} (\bibinfo{year}{2009}).

\bibitem[{\citenamefont{Schafer et~al.}(1998)}]{csmgr1998}
\bibinfo{author}{\bibfnamefont{C.} \bibnamefont{Sch$\ddot{a}$fer}}, 
  \bibinfo{author}{\bibfnamefont{M.~G.} \bibnamefont{Rosenblum}}, 
  \bibinfo{author}{ \bibnamefont{J.} \bibnamefont{Kurths}}, 
\bibnamefont{and} \bibinfo{author}{ \bibnamefont{H-H.} \bibnamefont{Abel}}, 
  \bibinfo{journal}{Nature(London)} \textbf{\bibinfo{volume}{392}},
  \bibinfo{pages}{239} (\bibinfo{year}{1998}).

\bibitem[{\citenamefont{Stefanovska et~al.}(2000)}]{ashh2000}
\bibinfo{author}{\bibfnamefont{A.} \bibnamefont{Stefanovska}}, 
  \bibinfo{author}{\bibfnamefont{H.} \bibnamefont{Haken}}, 
  \bibinfo{author}{\bibfnamefont{P. V. E.} \bibnamefont{McClintock}},
  \bibinfo{author}{\bibfnamefont{M.} \bibnamefont{Hozic}}, 
  \bibinfo{author}{\bibfnamefont{F.} \bibnamefont{Bajrovic}}, 
\bibnamefont{and}  \bibinfo{author}{\bibfnamefont{S.} \bibnamefont{Ribaric}},
  \bibinfo{journal}{Phys. Rev. Lett.} \textbf{\bibinfo{volume}{85}},
  \bibinfo{pages}{4831} (\bibinfo{year}{2000}).

\bibitem[{\citenamefont{Bartsch et~al.}(2007)}]{rbjwk2007}
\bibinfo{author}{\bibfnamefont{R.} \bibnamefont{Bartsch}}, 
  \bibinfo{author}{\bibfnamefont{J.~W.} \bibnamefont{Kantelhardt}}, 
  \bibinfo{author}{\bibfnamefont{T.} \bibnamefont{Penzel}}, 
\bibnamefont{and}  \bibinfo{author}{\bibfnamefont{S.} \bibnamefont{Havlin}},
  \bibinfo{journal}{Phys. Rev. Lett.} \textbf{\bibinfo{volume}{98}},
  \bibinfo{pages}{054102} (\bibinfo{year}{2007}).

\bibitem[{\citenamefont{Varela et~al.}(2001}]{fvjpl2001}
\bibinfo{author}{\bibfnamefont{F.} \bibnamefont{Varela}}, 
  \bibinfo{author}{\bibfnamefont{J. P.} \bibnamefont{Lachaux}},
 \bibinfo{author}{\bibfnamefont{E.} \bibnamefont{Rodriguez}}, 
\bibnamefont{and} \bibinfo{author}{ \bibnamefont{J.} \bibnamefont{Martinerie}}, 
  \bibinfo{journal}{Neuroscience (Nature Reviews)} \textbf{\bibinfo{volume}{2}},
  \bibinfo{pages}{229} (\bibinfo{year}{2001}).

\bibitem[{\citenamefont{Lian et~al.}(2004)}]{jljs2004}
\bibinfo{author}{\bibfnamefont{J.} \bibnamefont{Lian}}, 
  \bibinfo{author}{\bibfnamefont{J.} \bibnamefont{Shuai}}, 
\bibnamefont{and} \bibinfo{author}{ \bibnamefont{D. M.} \bibnamefont{Durand}}, 
  \bibinfo{journal}{J. Neural Eng.} \textbf{\bibinfo{volume}{1}},
  \bibinfo{pages}{46} (\bibinfo{year}{2004}).

\bibitem[{\citenamefont{Tass et~al.}(1998)}]{ptmgr1998}
\bibinfo{author}{\bibfnamefont{P.} \bibnamefont{Tass}}, 
  \bibinfo{author}{\bibfnamefont{M.~G.} \bibnamefont{Rosenblum}}, 
  \bibinfo{author}{\bibfnamefont{J.} \bibnamefont{Weule}}, 
  \bibinfo{author}{\bibfnamefont{J.} \bibnamefont{Kurths}}, 
  \bibinfo{author}{\bibfnamefont{A.} \bibnamefont{Pikovsky}}, 
  \bibinfo{author}{\bibfnamefont{J.} \bibnamefont{Volkmann}}, 
  \bibinfo{author}{\bibfnamefont{A.} \bibnamefont{Schnitzler}},
\bibnamefont{and}  \bibinfo{author}{\bibfnamefont{H.~-J.} \bibnamefont{Freund}},
  \bibinfo{journal}{Phys. Rev. Lett.} \textbf{\bibinfo{volume}{81}},
  \bibinfo{pages}{3291} (\bibinfo{year}{1998}).

\bibitem[{\citenamefont{Blasius et~al.}(1999)}]{bbah1990}
\bibinfo{author}{\bibfnamefont{B.} \bibnamefont{Blasius}}, 
  \bibinfo{author}{\bibfnamefont{A.} \bibnamefont{Huppert}}, 
\bibnamefont{and} \bibinfo{author}{ \bibnamefont{L.} \bibnamefont{Stone}}, 
  \bibinfo{journal}{Nature} \textbf{\bibinfo{volume}{399}},
  \bibinfo{pages}{354} (\bibinfo{year}{1999}).

\bibitem[{\citenamefont{Sismondo et~al.}(1990)}]{es1990}
\bibinfo{author}{\bibfnamefont{E.} \bibnamefont{Sismondo}}, 
  \bibinfo{journal}{Science} \textbf{\bibinfo{volume}{249}},
  \bibinfo{pages}{55} (\bibinfo{year}{1990}).

\bibitem[{\citenamefont{Amritkar et~al.}(2006)}]{reagr2006}
\bibinfo{author}{\bibfnamefont{R. E.} \bibnamefont{Amritkar}}, 
\bibnamefont{and}  \bibinfo{author}{\bibfnamefont{G.} \bibnamefont{Rangarajan}}, 
  \bibinfo{journal}{Phys. Rev. Lett.} \textbf{\bibinfo{volume}{96}},
  \bibinfo{pages}{258102} (\bibinfo{year}{2006}).

\bibitem[{\citenamefont{Rybski et~al.}(2003)}]{drsh2003}
\bibinfo{author}{\bibfnamefont{D.} \bibnamefont{Rybski}}, 
  \bibinfo{author}{\bibfnamefont{S.} \bibnamefont{Harlin}}, 
\bibnamefont{and}  \bibinfo{author}{\bibfnamefont{A.} \bibnamefont{Bunde}}, 
  \bibinfo{journal}{Physica A} \textbf{\bibinfo{volume}{320}},
  \bibinfo{pages}{601} (\bibinfo{year}{2003}).

\bibitem[{\citenamefont{Yamasaki}(2009)}]{kyag2009}
\bibinfo{author}{\bibfnamefont{K.} \bibnamefont{Yamasaki}}, 
  \bibinfo{author}{\bibfnamefont{A.} \bibnamefont{Gozolchiani}},
\bibnamefont{and} \bibinfo{author}{ \bibnamefont{S.} \bibnamefont{Harlin}}, 
  \bibinfo{journal}{Prog. Theor. Phys.} \textbf{\bibinfo{volume}{179}},
  \bibinfo{pages}{178} (\bibinfo{year}{2009}).

\bibitem[{\citenamefont{Maraun}(2005)}]{dmjk2005}
\bibinfo{author}{\bibfnamefont{D.} \bibnamefont{Maraun}}, 
\bibnamefont{and} \bibinfo{author}{ \bibnamefont{J.} \bibnamefont{Kurths}}, 
  \bibinfo{journal}{Geophys. Res. Lett.} \textbf{\bibinfo{volume}{32}},
  \bibinfo{pages}{L15709} (\bibinfo{year}{2005}).

\bibitem[{\citenamefont{Senthilkumar et~al.}(2006)}]{dvskml2006}
\bibinfo{author}{\bibfnamefont{D.~V.} \bibnamefont{Senthilkumar}}, 
  \bibinfo{author}{\bibfnamefont{M.} \bibnamefont{Lakshmanan}}, 
\bibnamefont{and}  \bibinfo{author}{ \bibnamefont{J.} \bibnamefont{Kurths}}, 
  \bibinfo{journal}{Phys. Rev. E} \textbf{\bibinfo{volume}{74}},
  \bibinfo{pages}{035205(R)} (\bibinfo{year}{2006}).

\bibitem[{\citenamefont{Senthilkumar et~al.}(2008)}]{dvskml2008}
\bibinfo{author}{\bibfnamefont{D.~V.} \bibnamefont{Senthilkumar}}, 
  \bibinfo{author}{\bibfnamefont{M.} \bibnamefont{Lakshmanan}}, 
\bibnamefont{and}  \bibinfo{author}{ \bibnamefont{J.} \bibnamefont{Kurths}}, 
  \bibinfo{journal}{Chaos} \textbf{\bibinfo{volume}{18}},
  \bibinfo{pages}{023118} (\bibinfo{year}{2008}).

\bibitem[{\citenamefont{Romano et~al.}(2005)}]{mcrmt2005}
\bibinfo{author}{\bibfnamefont{M.~C.} \bibnamefont{Romano}}, 
  \bibinfo{author}{\bibfnamefont{M.} \bibnamefont{Thiel}}, 
  \bibinfo{author}{\bibfnamefont{J.} \bibnamefont{Kurths}}, 
  \bibinfo{author}{\bibfnamefont{I.~Z.} \bibnamefont{Kiss}}, 
\bibnamefont{and} \bibinfo{author}{\bibfnamefont{J.~L.} \bibnamefont{Hudson}}, 
  \bibinfo{journal}{Europhys. Lett.} \textbf{\bibinfo{volume}{71}},
  \bibinfo{pages}{466} (\bibinfo{year}{2005}).

\bibitem[{\citenamefont{Marwan et~al.}(2007)}]{nmmcr2007}
\bibinfo{author}{\bibfnamefont{N.} \bibnamefont{Marwan}},
\bibinfo{author}{\bibfnamefont{M. C.} \bibnamefont{Romano}},
\bibinfo{author}{\bibfnamefont{M.} \bibnamefont{Thiel}},
  \bibnamefont{and}
  \bibinfo{author}{\bibfnamefont{J.} \bibnamefont{Kurths}},
  \bibinfo{journal}{Phys. Rep.} \textbf{\bibinfo{volume}{438}},
  \bibinfo{pages}{237} (\bibinfo{year}{2007}). 

\bibitem[{\citenamefont{Pereira et~al.}(2007)}]{tpmsb2007}
\bibinfo{author}{\bibfnamefont{T.} \bibnamefont{Pereira}}, 
  \bibinfo{author}{\bibfnamefont{M.~S.} \bibnamefont{Baptista}}, 
\bibnamefont{and} \bibinfo{author}{\bibfnamefont{J.} \bibnamefont{Kurths}}, 
  \bibinfo{journal}{Phys. Rev. E} \textbf{\bibinfo{volume}{75}},
  \bibinfo{pages}{026216} (\bibinfo{year}{2007}).

\bibitem[{\citenamefont{Chil et~al.}(2002)}]{chil2002}
\bibinfo{author}{\bibfnamefont{Chil-Min} \bibnamefont{Kim}}, 
  \bibinfo{author}{\bibfnamefont{S.} \bibnamefont{Rim}}, 
\bibnamefont{and} \bibinfo{author}{\bibfnamefont{W-H.} \bibnamefont{Kye}}, 
  \bibinfo{journal}{Phys. Rev. Lett.} \textbf{\bibinfo{volume}{88}},
  \bibinfo{pages}{014103} (\bibinfo{year}{2001}).

\bibitem[{\citenamefont{Keneko}(1990)}]{kk1990}
\bibinfo{author}{\bibfnamefont{K.} \bibnamefont{Kaneko}}, 
  \bibinfo{journal}{Physica D} \textbf{\bibinfo{volume}{41}},
  \bibinfo{pages}{137} (\bibinfo{year}{1990}).

\bibitem[{\citenamefont{Sherman}(1994)}]{as1994}
\bibinfo{author}{\bibfnamefont{A.} \bibnamefont{Sherman}}, 
  \bibinfo{journal}{Bull. Math. Biol.} \textbf{\bibinfo{volume}{56}},
  \bibinfo{pages}{811} (\bibinfo{year}{1994}).

\bibitem[{\citenamefont{Strogatz et~al.}(1993)}]{shs1993}
\bibinfo{author}{\bibfnamefont{S.~H.} \bibnamefont{Strogatz}}, 
\bibnamefont{and} \bibinfo{author}{\bibfnamefont{I.} \bibnamefont{Stewart}}, 
  \bibinfo{journal}{Sci. Am.} \textbf{\bibinfo{volume}{269}},
  \bibinfo{pages}{102} (\bibinfo{year}{1993}).

\bibitem[{\citenamefont{Terry et~al.}(1999)}]{jrt1999}
\bibinfo{author}{\bibfnamefont{J.~R.} \bibnamefont{Terry}}, 
  \bibinfo{author}{\bibfnamefont{K.~S.} \bibnamefont{Thornburg}}, 
  \bibinfo{author}{\bibfnamefont{Jr.,~D.~J.} \bibnamefont{DeShazer}}, 
  \bibinfo{author}{\bibfnamefont{G.~D.} \bibnamefont{Vanwiggeren}}, 
  \bibinfo{author}{\bibfnamefont{S.} \bibnamefont{Zhu}}, 
  \bibinfo{author}{\bibfnamefont{P.} \bibnamefont{Ashwin}},
\bibnamefont{and} \bibinfo{author}{\bibfnamefont{R.} \bibnamefont{Roy}}, 
  \bibinfo{journal}{Phys. Rev. E} \textbf{\bibinfo{volume}{59}},
  \bibinfo{pages}{4036} (\bibinfo{year}{1999}).

\bibitem[{\citenamefont{Mackey et~al.}(1977)}]{mcmlg1977}
\bibinfo{author}{\bibfnamefont{M.~C.} \bibnamefont{Mackey}}, 
\bibnamefont{and}  \bibinfo{author}{\bibfnamefont{L.} \bibnamefont{Glass}},
  \bibinfo{journal}{Science} \textbf{\bibinfo{volume}{197}},
  \bibinfo{pages}{287} (\bibinfo{year}{1977}).

\bibitem[{\citenamefont{Zhou et~al.}(1999)}]{cschl1999}
\bibinfo{author}{\bibfnamefont{C.} \bibnamefont{Zhou}}, 
\bibnamefont{and}  \bibinfo{author}{\bibfnamefont{C.~H.} \bibnamefont{Lai}},
  \bibinfo{journal}{Phys. Rev. E} \textbf{\bibinfo{volume}{60}},
  \bibinfo{pages}{320} (\bibinfo{year}{1999}).

\bibitem[{\citenamefont{Pyragas}(1998)}]{kp1998}
\bibinfo{author}{\bibfnamefont{K.} \bibnamefont{Pyragas}}, 
  \bibinfo{journal}{Phys. Rev. E} \textbf{\bibinfo{volume}{58}},
  \bibinfo{pages}{3067} (\bibinfo{year}{1998}).

\bibitem[{\citenamefont{szendro et~al.}(2005)}]{igsjml2005}
\bibinfo{author}{\bibfnamefont{I.~G.} \bibnamefont{Szendro}}, 
\bibnamefont{and}  \bibinfo{author}{\bibfnamefont{J.~M.} \bibnamefont{Lopez}},
  \bibinfo{journal}{Phys. Rev. E} \textbf{\bibinfo{volume}{71}},
  \bibinfo{pages}{055203} (\bibinfo{year}{2005}).

\bibitem[{\citenamefont{Namajunas et~al.}(1995)}]{ankp1995}
\bibinfo{author}{\bibfnamefont{A.} \bibnamefont{Namaj\={u}nas}}, 
\bibnamefont{and}  \bibinfo{author}{\bibfnamefont{K.} \bibnamefont{Pyragas}},
\bibnamefont{and}  \bibinfo{author}{\bibfnamefont{A.} \bibnamefont{Tama\v{s}evi\v{e}ins}},
  \bibinfo{journal}{Phys. Lett. A} \textbf{\bibinfo{volume}{201}},
  \bibinfo{pages}{42} (\bibinfo{year}{1995}).

\bibitem[{\citenamefont{Osipov}(1986)}]{gvocsz2007}
\bibinfo{author}{\bibfnamefont{G.~V.}~\bibnamefont{Osipov}},
\bibinfo{author}{\bibfnamefont{C.}~\bibnamefont{Zhou}}, \bibnamefont{and}
  \bibinfo{author}{\bibfnamefont{J.}~\bibnamefont{Kurths}},
  \emph{\bibinfo{title}{Synchronization in Oscillatory Networks
 }} (\bibinfo{publisher}{Springer},
  \bibinfo{address}{Berlin}, \bibinfo{year}{2007}).

\bibitem[{\citenamefont{Senthilkumar lakshmanan}(2008)}]{dvskmlejp2008}
\bibinfo{author}{\bibfnamefont{D.~V.} \bibnamefont{Senthilkumar}}, 
  \bibinfo{author}{\bibfnamefont{M.} \bibnamefont{Lakshmanan}}, 
\bibnamefont{and}  \bibinfo{author}{ \bibnamefont{J.} \bibnamefont{Kurths}}, 
  \bibinfo{journal}{Eur. Phys. J. Special Topics} \textbf{\bibinfo{volume}{164}},
  \bibinfo{pages}{35} (\bibinfo{year}{2008}).

\bibitem[{\citenamefont{Moreno}(2004)}]{ymafp2004}
\bibinfo{author}{\bibfnamefont{Y.} \bibnamefont{Moreno}}, 
\bibnamefont{and}  \bibinfo{author}{ \bibnamefont{A. F.} \bibnamefont{Pacheco}}, 
  \bibinfo{journal}{Europhys. Lett.} \textbf{\bibinfo{volume}{68}},
  \bibinfo{pages}{603} (\bibinfo{year}{2004}).
%%
\end{thebibliography}
\end{document}